\documentclass[12pt,aps,tightenlines,superscriptaddress]{revtex4}
\usepackage{graphicx}

\begin{document}

\title{
Search for and study of $\eta$-mesic nuclei
in $pA$-collisions at the JINR LHE nuclotron%
\footnote{Supported by the RFBR grants 02-02-16519 and 03-02-17376.}}

\author{M.Kh.~Anikina}  
\author{Yu.S.~Anisimov}
\author{A.S.~Artemov}
\author{S.V.~Afanasev}
\author{D.K.~Dryablov}
\author{V.I.~Ivanov}
\author{V.A.~Krasnov}
\author{S.N.~Kuznetsov}
\author{A.N.~Livanov}
\author{A.I.~Malakhov}
\author{P.V.~Rukoyatkin}
\affiliation{Joint Institute for Nuclear Research, LHE, Dubna}

\author{V.A.~Baskov}
\author{A.I.~Lebedev}
\author{A.I.~L'vov}
\author{L.N.~Pavlyuchenko}
\author{V.P.~Pavlyuchenko}
\author{V.V.~Polyansky}
\author{S.S.~Sidorin}
\author{G.A.~Sokol}
\author{E.I.~Tamm}
\affiliation{Lebedev Physical Institute, Moscow}

\author{E.V.~Balandina}
\author{E.M.~Leikin}
\author{N.P.~Yudin}
\affiliation{Institute of Nuclear Physics, MSU, Moscow}

\author{Yu.N.~Uzikov}
\affiliation{Joint Institute for Nuclear Research, LNP, Dubna}

\author{V.B.~Belyaev}
\author{N.V.~Shevchenko}
\affiliation{Joint Institute for Nuclear Research, LTP, Dubna}

\author{V.Yu.~Grishina}
\author{A.B.~Kurepin}
\affiliation{Institute for Nuclear Research, Moscow}

\author{L.A.~Kondratyuk}
\affiliation{Institute of Theoretical and Experimental Physics, Moscow}

\author{L.M.~Lazarev}
\affiliation{Institute of Experimental Physics, Sarov}

\author{A.V.~Kobushkin}
\affiliation{Institute of Theoretical Physics, Kiev}

\author{S.~Gmutsa}
\author{M.~Morha\v c}
\affiliation{Institute of Physics, Bratislava}

\author{S.~Wycech}
\affiliation{Institute of Nuclear Physics, Warsaw\vspace*{1em}}


\begin{abstract}
An approved experiment at the internal proton beam of the JINR nuclotron
on a search for $\eta$-mesic nuclei in the reaction
$pA\to np + {}_\eta(A-1)\to np+\pi^-p+X$ is briefly presented.
\end{abstract}

\maketitle

\subsection*{Introduction}

The present project is aimed at a search for and study of hadronic systems
of a new kind, $\eta$-mesic nuclei, which are bound states of the
$\eta$-meson and a nucleus. Such states were predicted long ago
\cite{haider86} after noticing that an interaction between a slow
$\eta$-meson and the nuclear matter is attractive. At zero energy, an
optical $\eta A$-potential is equal to $V(r)= -(2\pi\rho(r)/\mu)a_{\eta N}$,
where $\mu$ is the reduced mass of $\eta$ and $N$, $\rho(r)$ is the nuclear
density, and $a_{\eta N}$ is the $\eta N$-scattering length. Estimates
of this scattering length are derived from analyses of 
$\pi N\to\pi N$ and $\pi N\to\eta N$ reactions (e.g. $a_{\eta N} =
(0.75-i0.27)~\rm fm$ according to \cite{green97}). The positive sign of
${\rm Re}\,a_{\eta N}$ just means an attractive $\eta A$ optical potential.

Physics of $\eta$-mesic nuclei is a new field in nuclear and particle
physics. A prominent feature of the $\eta N$ interaction is the strong
inelastic mode $\eta N\to \pi N$ and a formation of the near-threshold
$S_{11}(1535)$ resonance. The life-time of the $\eta$-nucleus is much
shorter than the life-time of $\eta$ itself, and the wave function of an
$\eta$-nucleus ${}_\eta A$ is not a pure $\eta A$ state but rather a
superposition involving the pure state, the $S_{11}$-resonance replacing a
nucleon in the nucleus $A$, and a pion continuum.

Investigations of $\eta$-nuclei open new possibilities for studying the
$\eta N$ interaction. All data on $\eta N$ scattering are indirect and
obtained from analyses of reactions in which $\eta N$ is produced in the
final state. Results of such analyses suffer from large theoretical
uncertainties. Depending on assumptions made, various results for
$a_{\eta N}$ are derived in literature which are different by the factor
of 4. On the other hand, the average $\eta A$ potential in $\eta$-nuclei is
mainly formed by $\eta N$-scattering. Knowning the $\eta A$-potential or,
at least, binding energies, one can learn more on the underlying $\eta N$
scattering.

One of aims in studies of $\eta$-nucleus systems is probing the
$S_{11}(1535)$ resonance in the nuclear matter. Both the mass and the width
of this resonance are changed \cite{jido02} --- in particular, because new
decay channels are opened in the nuclear matter, such as the collisional
decay $S_{11}+N\to N+N$. Hadron masses are known to be closely related with
the chiral condensate $q\bar q$. Understanding of the modification of the
chiral condensate in the nuclear matther is an important and challenging
problem. A presence of strange quarks in the $\eta$-meson allows to probe
the strange condensate $s\bar s$ in addition to nonstrange condensates
$u\bar u$ and $d\bar d$.

The first experiment on a search for $\eta$-mesic nuclei has been performed
in Brookhaven \cite{chrien88}. Narrow peaks of a few-MeV width in the
missing-mass spectrum of $\pi^+ A\to p X$ were then expected. However,
no clean signal was found. Later it became clear that the peaks have not to
be so narrow and that a better strategy of searching for $\eta$-nuclei is
required. Instead of doing missing-mass experiments, it was proposed to
detect decay products of $\eta$-nuclei \cite{sokol91}. Following this idea,
an experiment was performed at the photon beam of the LPI electron
synchroton and for the first time a direct signal of formation of
$\eta$-nuclei had been found \cite{sokol99,sokol00}.

Last years further experiments on searching for $\eta$-nuclei are on their
way -- at the ion beam of GSI (Darmstadt) \cite{hayano98} and at the proton
beam of COSY (Yulich) \cite{gilitzer01}. A possibility to use the photon
beam in Grenoble is also considered \cite{baskov03}. Very recently, a
collaboration working at the electron microtron MAMI-B (Mainz) observed a
bound state of the $\eta$-meson in ${}^3$He \cite{pfeiffer04}.

\begin{figure}[hbt]
\centering{
\includegraphics[height=3.5cm]{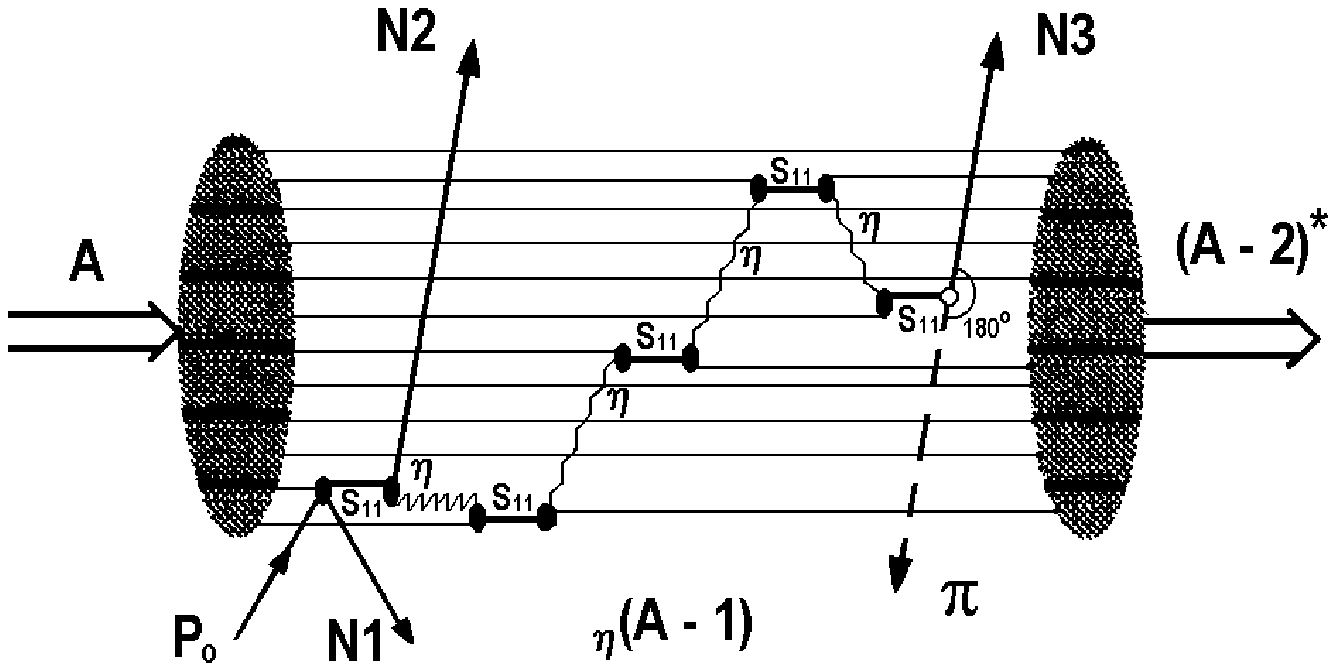}
\includegraphics[height=3.5cm]{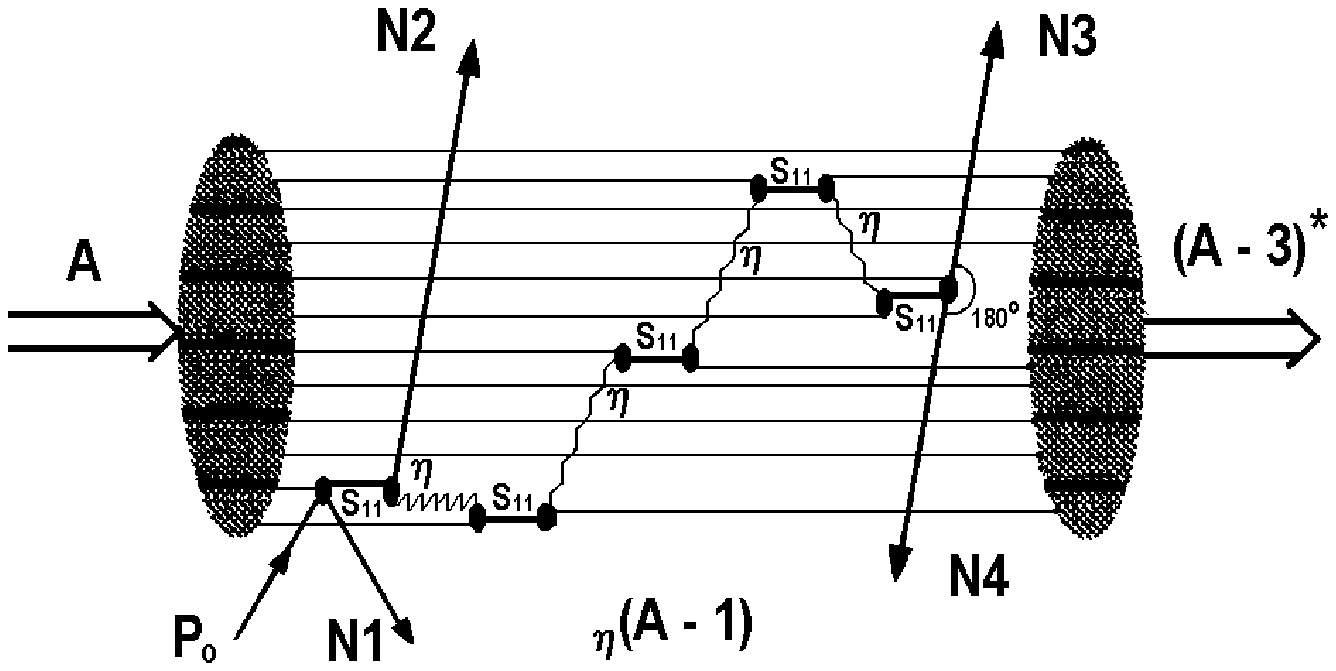}}
\caption{Formation, evolution and decay of a $\eta$-nucleus  
in a $pA$-collision. Two decay modes are shown: $\pi N$ (probability 90\%)
and $NN$ (probability 10\%).}
\label{fig1}
\end{figure}

The present project is aimed at searching for $\eta$-nuclei at the internal
proton beam of the JINR LHE nuclotron in reactions
\begin{eqnarray}
  p + A \to n + p + {}_{\eta}(A - 1) &\to& n + p + \pi^- + p + X, \\
  p + A \to n + p + {}_{\eta}(A - 1) &\to& n + p + p + p + X.
\end{eqnarray}
An incident proton of the energy $\sim 2$~GeV hits a neutron in the nuclear
target and produces a slow $\eta$-meson through the subprocess $p+n\to
n+p+\eta$ yielding also two nucleons flying forward. These nucleons tag the
created $\eta$-meson and they are included into the trigger. The slow
$\eta$-meson is captured by the nucleus and forms an $\eta$-nucleus that
then decays producing a $\pi^-p$ or $pp$ pair (see Fig.~\ref{fig1}). A
bound state of $\eta$ is expected to be seen as a peak in the energy
spectrum of these pairs. Thus, predictions of numerious calculations for
binding energies and widths can be tested. Moreover, relative probabilities
of the $\pi N$ and $NN$ decay modes of $\eta$-nuclei can be measured.

\subsection*{Kinematics of $\eta$-nucleus formation and decay}

Eta-nuclei are highly unstable. Their widths are typically
$\sim20{-}30$ MeV. When a slow $\eta$ travels in a nucleus, multiple
$\eta N$-scattering occurs through excitations of intermediate
$S_{11}(1535)$ resonances at different nucleons:
$\eta + N_1 \to S_{11} \to \eta + N_1$,
$\eta + N_2 \to S_{11} \to \eta + N_2$, \ldots
$\eta + N_k \to S_{11} \to \pi  + N_k$.
The last step here is a conversion of $\eta$ into an energetic $\pi N$
pair which escapes the nucleus. The final $\pi N$ pair carries the
kinetic energy of about $m_\eta-m_\pi\simeq 400$ MeV and zero 3-momentum
(up to Fermi-motion effects). That kinetic energy is shared between $\pi$
and $N$ as $T_\pi\simeq 310$ MeV and $T_N\simeq 90$ MeV. 3-momenta of the
pion and nucleon are nearly opposite, $\theta_{\pi N}\simeq 180^\circ$.
In the case of the $S_{11}N\to NN$ mode of the $\eta$-nucleus decay,
nucleons in the $NN$ pair have energies of about $m_\eta/2 = 270$~MeV and
again nearly opposite 3-momenta.

It is essential that $\pi N$ pairs with $\theta_{\pi N}\simeq 180^\circ$
and flying transversely to the beam cannot be easily produced in an
alternative way. For example, pairs created in the process
$N+N\to N+S_{11} \to N+\pi+N$ carry a large total 3-momentum needed
to produce the $S_{11}$, and they have $\theta_{\pi N} \approx 100^\circ$.
A thermalization or capture of $\eta$ is really needed to create transverse pairs.

Incident protons produce $\eta$-mesons through a pion exchange as shown in
Fig.~2. Due to isotopic factors and antisimmetrization, $\eta$ is mainly
produced in a $pn$ (rather than $pp$) charged-exchange reaction with
emission of a leading forward neutron. This neutron has a low transverse
momentum $\lesssim 200{-}300$ MeV/c and, at energies of our interest
($T_{\rm beam}\sim2$~GeV), travels at the angle $6^\circ{-}9^\circ$ with
respect to the beam. The proton produced together with a slow eta, has the
momentum of about 800 MeV/c (or kinetic energy of 300 MeV). Its typical
angle is $\sim 20{-}30^\circ$.

In the case of the carbon target, the total cross section of $\eta$-nucleus
formation is expected to be $\sigma({}_{\eta}A) \sim 10~\mu b$,
or 1\% of the total $\eta$-production cross section. Meanwhile
the inclusive cross section of inelastic processes in $p\,$C collisions is
much bigger ($\simeq 250$~mb) thus leading to dangerous backgrounds.
That is why it is absolutely necessary to include into the trigger
both the $\pi N$ pair and the forward-flying nucleon(s).
In the proposed experiment isotopic modes with charged particles will be
detected: the transverse $\pi^- p$ or $pp$ pair from the $\eta$-nucleus
decay, the proton $p_1$ from the formation stage, and optionally the
forward-flying neutron.

\begin{figure}[hbt]
\centering{
\includegraphics[width=8cm]{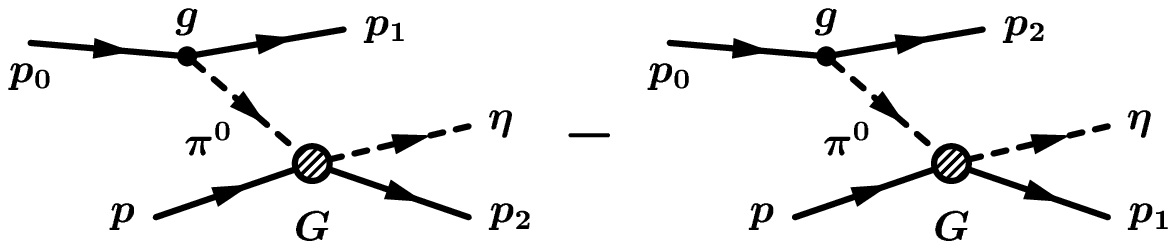} \hspace*{0em}
\includegraphics[width=8cm]{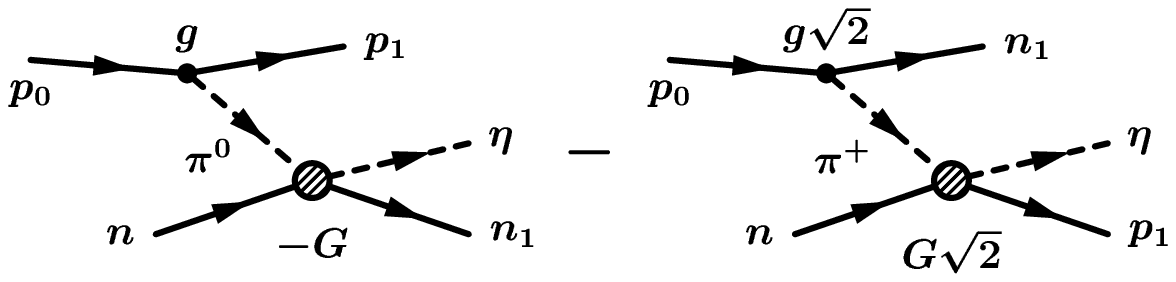}}
\caption{A mechanism of slow-$\eta$ production in $pp$ and $pn$ collisions.
The blob means an excitation of an intermediate $S_{11}(1535)$ resonance.}
\label{fig2}
\end{figure}

\subsection*{Experimental setup}                

The planned experimental setup (Fig.~\ref{fig3}) includes a two-arm
spectrometer for detecting $\pi^-p$ and $pp$ pairs from $\eta$-nucleus
decays and ring spectrometers for detecting $p$ and $n$ emitted at the
first stage of the reactions (1),~(2). The setup will include the following
systems:

\begin{figure}[h]
\includegraphics[width=6cm]{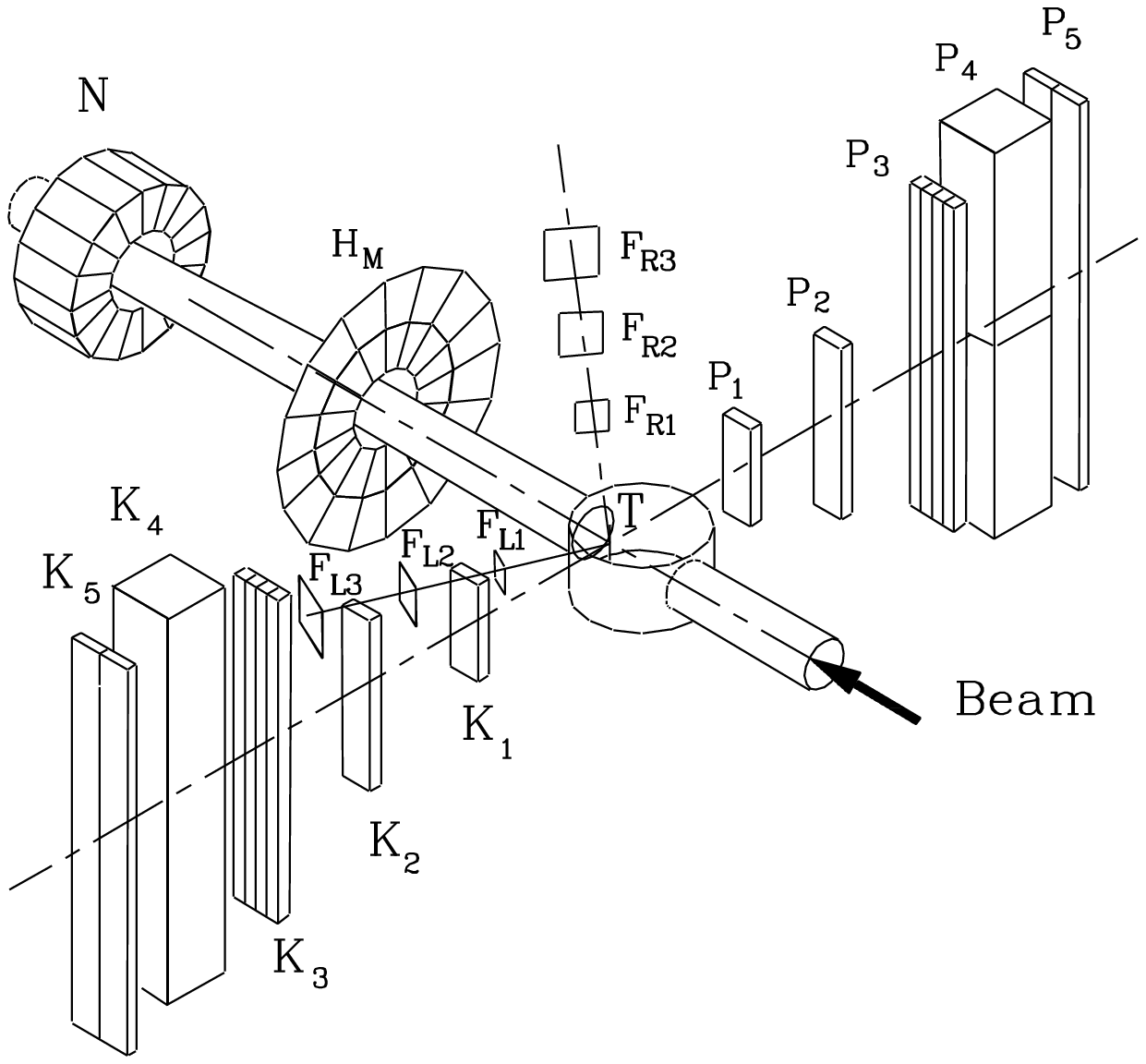}
\hspace*{4em}
\raisebox{2em}{\includegraphics[width=6cm]{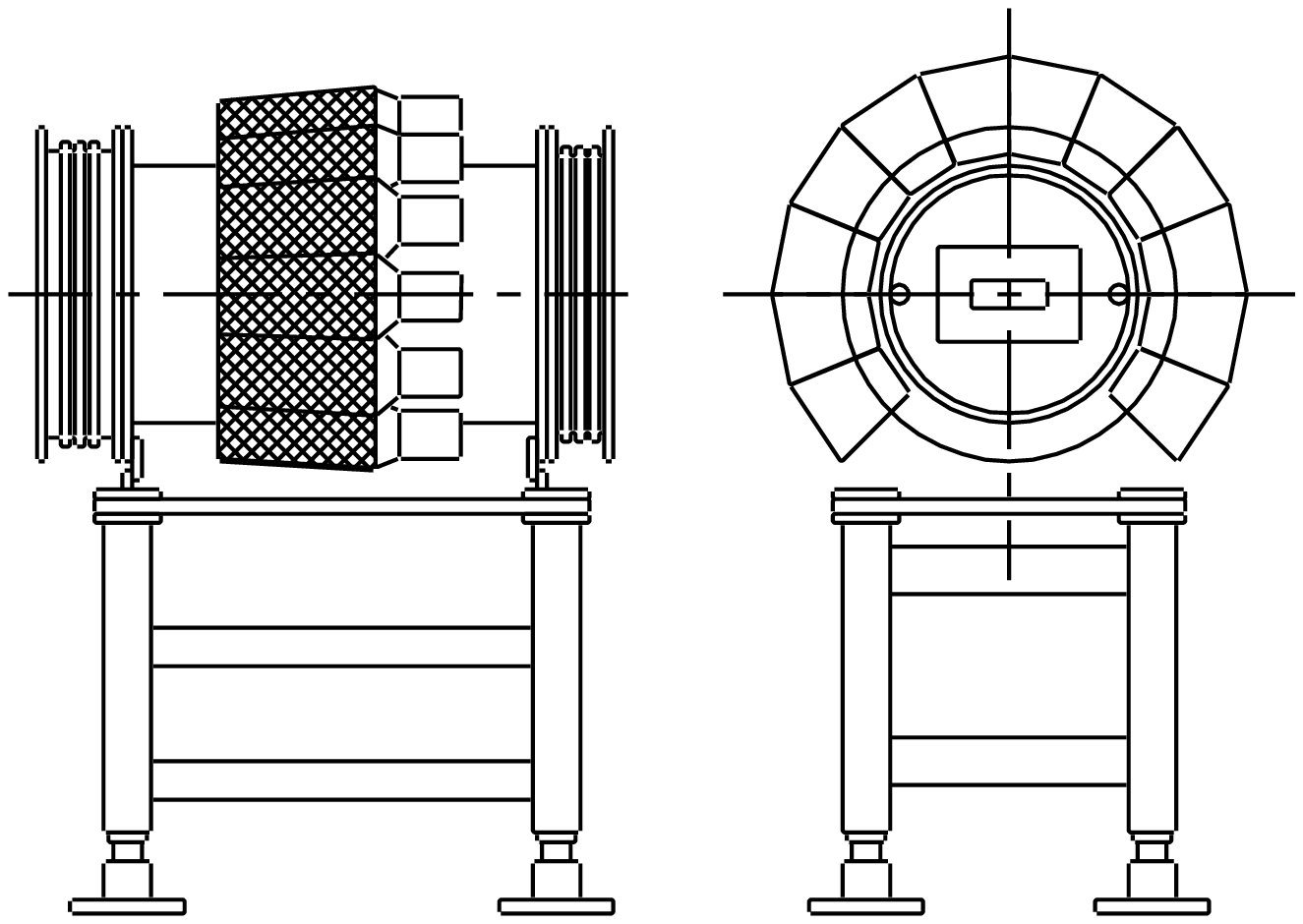}}
\caption{Layout of the experimental setup and the neutron detector.} 
\label{fig3}
\end{figure}

1) A monitor system consisting of 4 scintillator telescopes ---
two triple monitors $(F_L,~F_R$) and two double monitors $(B_L,~B_R$).

2) A 32-channel scintillator hodoscope $H_M$ for detection of
forward-flying protons.

3) A two-arm spectrometer that serves for detecting $\pi^- p$ and $pp$
pairs and measuring particle momenta (time-of-flight) and kinetic energies
($dE/dx$). One arm of this spectrometer already exists: that is the
spectrometer SCAN \cite{SCAN}. Its time-of-flight part includes
scintillator detectors P1 and P3 (P3 consists of 4 segments) and a
solid-state threshold Cherenkov counter P2. The $dE/dx$ part consists of
two thick scintillators P4 and two counters P5 which select high-energy
particles passing through all layers. Each of these detectors has two
phototubes at top and bottom edges. The second arm of the spectrometer is
yet to be built.

4) A neutron detector (Fig.~3) will be built and installed beyond the
vacuum tube at $\sim 3$~m from the target. That will be a segmented ring of
large-volume scintillator covered with veto counters excluding charged
particles. A cryogenic shell of the accelerator serves as an absorber of
charged particles that reduces a overall load of the detector.

In order to select usefull events from the expected flow of $10^4$
triggers/s, a fast two-level trigger will be used. At the first level,
events with transverse $180^\circ$ pairs will be selected. Fast
identification of particles will be performed using the Cherenkov
counter. At the second level, the selected events will be checked using
information from the proton and neutron detectors.

The experiment will be performed at the internal proton beam of the JINR
nuclotron. Thin filament (or film) targets of various $A$ (Be, C, Al) and
thickness $10{-}20~\mu$m will be used. Targets are placed into a target
block which enables one to remotely move targets transversely to the beam
having the size $\sigma_b=0.75$~mm. This allows to maintain a constant
collision rate during $\sim$5~s and reduce an average load of detectors what is
important for coincidence experiments. Multiple passes of proton bunches
through the target ($6 \cdot 10^6$ times during 5~s) result in a
sufficient luminosity of the setup: up to $L \simeq 0.5\cdot 10^{32}~\rm
cm^{-2}~s^{-1}$ in average per one accelerator cycle (10~s) with $10^9$
initially accelerated  protons. There is practically no background caused
by secondary interactions of particles produced in a very thin target
with the target material.

\subsection*{Signal and background}

Expected count rates of the setup for signal and background
are summarized in the Table~1 for several variants of coincidences
with a $10~\mu$m carbon filament target and $10^9$ initial protons
at the accelerator orbit.

\begin{table}[h]
\caption{Signal and random-coincidence background rates for different
variants of coincidences at the full luminosity of $10^9$ initial protons
at the accelerator orbit and a $10~\mu$m carbon filament target.}
\begin{center}
\begin{tabular}{|llll|c|c|}
\hline
\multicolumn{4}{|c|}{coincidences} & signal & background \\
 &&&& events/hour & events/hour \\
\hline
$Y (p\pi^-)$      & $\langle\theta_{p\pi}\rangle = 180^\circ$
  &&& 1400 & 6500 \\
\hline
$Y (pp)$          & $\langle\theta_{pp}\rangle = 180^\circ$
  &&& 230 & \\
\hline
$Y (p\pi^-,p_1)$  & $\langle\theta_{p\pi}\rangle = 180^\circ$
  & $\theta_{p_1} = (15{-}20)^\circ$  && 250 & 960 \\
\hline
$Y (pp,p_1)$      & $\langle\theta_{pp}\rangle = 180^\circ$
  & $\theta_{p_1} = (15{-}20)^\circ$ && 40 & \\
\hline
$Y(p\pi^-,p_1n)$  & $\langle\theta_{p\pi}\rangle = 180^\circ$
  & $\theta_{p_1} = (15{-}20)^\circ$ & $\theta_n =(7{-}11)^\circ$ & 35 & 1 \\
\hline
$Y(pp,p_1n)$      & $\langle\theta_{pp}\rangle = 180^\circ$
  & $\theta_{p_1} = (15{-}20)^\circ$ & $\theta_n =(7{-}11)^\circ$ & 6 & \\
\hline
\end{tabular}
\end{center}
\end{table}

Yields of the double-coincidence events $p\pi^-$ and $pp$ from the reaction
of formation of $\eta$-nuclei are determined by the expected cross section
$\sigma({}_\eta A)\simeq 10~\mu$b of $\eta$-nucleus formation in
$p\,$C-collisions, branching ratios ${\rm Br}(p\pi^-)\simeq 0.15$ and
${\rm Br}(pp)\simeq 0.025$ of the $\eta$-nucleus to decay through the
$p\pi^-$ and $pp$ modes, respectively, the solid angle $\Omega_\pi =
0.1~$sr of the two-arm spectrometer, and the geometric probability
$g_p\simeq 0.2$ of the proton to hit the second arm provided another
particle of the correlated pair hits the first arm (this probability is
determined by a theoretically expected width of the angular correlation).

Addition of the proton $p_1$ to the coincidences reduces count rates by the
geometric probability $g_{p_1}\simeq 0.18$ of the proton $p_1$ emitted in
the reaction of slow-$\eta$ production to fly in the angular range
$(15{-}20)^\circ$ covered by the hodoscope $H_M$. Addition of the neutron
further reduces count rates by the neutron detector efficienty
$\epsilon_n\simeq 0.3$ (as determined by GEANT simulation) and the
geometric probability $g_n\simeq 0.5$ of the neutron emitted in the same
reaction to fly in the angular range $(7{-}11)^\circ$ covered by the
neutron detector (this probability is determined by a simulation of 
processes shown in Fig. \ref{fig2}).

GEANT simulation of $3\cdot 10^8$ $p\,$C-collisions (ignoring the
possibility of $\eta$-nucleus formation) found no events with two charged
particles $\pi^\pm$ or $p$ flying at $\sim90^\circ$ and one proton flying
forward in the kinematical range of interest. Therefore backgrounds for
triple and 4-fold coincidences are only related with random coincidences.
Those backrounds were found as follows.

Using earlier Dubna data obtained with a propane chamber at $T_p=3.3~$GeV,
average loads of elements of the hodoscope $H_M$ positioned at 1.5~m from
the target and detectors of the two-arm spectrometer have been estimated
and found to be $< 5\cdot 10^5~\rm s^{-1}$ in conditions of the planned
experiment. Consistent results were also obtained via GEANT simulation
which also gave information on loads with neutral particles. Moreover, a
test run with carbon and polyethilene targets was done in December 2003 at
the JINR nuclotron and the SCAN spectrometer. In that run, the beam energy
was 1.5~GeV, 2~GeV, 3~GeV and 4.2~GeV and the beam intensity was
$I_p=(0.6{-}1)\cdot 10^{10}~\rm s^{-1}$. Loads of all detector elements
were measured and found to be less than $2\cdot 10^5~\rm s^{-1}$. Random
coincidences obtained with such loads and the time resolution of the
coincidence scheme, $\tau=20$~ns, are given in the Table~1.

At the full intensity created by $10^9$ protons, only 4-fold coincidences
clearly select signal events from the background, whereas the rate of
triple conicidences is bigger for the background than that for the signal.
Such an unfavorite situation can be cured by a reduction of the luminosity
--- via a reduction of the beam intensity or placing the target more away
from the beam axis. When the luminosity is decreased by $N=10$ times, the
rate of the signal is reduced by the factor of $N=10$ too, whereas the rate
of random 2-coincidences is reduced by $N^2=100$ and that of triple
coincidences is reduced by $N^3=1000$. Such a $N=10$ reduction of the
luminosity still makes it possible to obtain a sufficient number of events
with double and triple coincidences for a further analysis and
determination of energy distributions.

\subsection*{Conclusions}

The above estimates show that the main aims of the proposed experiment can be achieved.
Among them are:

-\quad search for $\eta$-mesic nuclei formed in $pA$ collisions and their
separation from background;

-\quad measurement of the total cross section $\sigma({}_{\eta} A)$ of
$\eta$-nucleus production in $pA$ collisions and its energy and
$A$-dependence;

-\quad measurement of the total energy distribution for $p\pi^-$ pairs
arising from decays of $\eta$-nuclei and determination of the energy level
and the width of the formed $\eta$-nucleus;

-\quad measurement of the branching ratios of $p\pi^-$ and $pp$ modes
of $\eta$-nucleus decays.

\makeatletter
\def\bibsection{%
 \@ifx@empty\refname{%
  \par
 }{%
  \let\@hangfroms@section\@hang@froms
  \subsection*{\refname}%
  \@nobreaktrue
 }%
}%
\makeatother


\begin{thebibliography}{99}

\bibitem{haider86}
  Q. Haider and L.C. Liu, Phys. Lett. B172 (1986) 257; Phys. Rev. C34 (1986) 1845.
\bibitem{green97}
  A.M. Green and S. Wycech, Phys. Rev. C55 (1997) R2167; nucl-th/9703009.

\bibitem{jido02}
  D. Jido et al., Phys. Rev. C66 (2002) 045202; nucl-th/0206043.

\bibitem{chrien88}
  R.E. Chrien et al., Phys. Rev. Lett. 60 (1988) 2595.
\bibitem{sokol91}
  G.A. Sokol and V.A. Tryasuchev,
  Kratkie Soobsh. Fiz. [English transl.: Sov. Phys. --
  Lebedev Institute Reports] Issue \#4 (1991) 23.
\bibitem{sokol99}
  G.A. Sokol et al., Fizika B8 (1999) 81; nucl-ex/9905006.
\bibitem{sokol00}
  G.A. Sokol et al., Pisma v EChaYa \#5 [102] (2000) 71.

\bibitem{hayano98}
  R.S. Hayano et al., nucl-th/9806012.
\bibitem{gilitzer01}
  A. Gilitzer, Proposal COSY-TOF collaboration, Oct. 2001.
\bibitem{baskov03}
  V.A. Baskov et al., nucl-ex/0306011.
\bibitem{pfeiffer04}
  M. Pfeiffer et al., Phys. Rev. Lett. 92 (2004) 252001; nucl-ex/0312011.

\bibitem{SCAN}
  S.V. Afanasev et al., JINR Rapid. Comm., 5[91]-98 (1988) 25.

\end{thebibliography}
\end{document}